\documentstyle[aps,epsf,floats]{revtex}

\tighten
\newcommand{\beq}{\begin{equation}}
\newcommand{\eeq}{\end{equation}}
\newcommand{\ba}{\begin{eqnarray}}
\newcommand{\ea}{\end{eqnarray}}
\newcommand{\nn}{\nonumber}
\newcommand{\dz}{\int \frac{d^{4}z}{(2\pi)^4}}
\newcommand{\dzp}{\frac{d^{4}z'}{(2\pi)^4}}

\newcommand{\bpt}{\bm p_T^{}}
\newcommand{\bkt}{\bm k_T^{}}
\newcommand{\bqt}{\bm q_T^{}}
\newcommand{\psibar}{\overline{\psi}}
\newcommand{\la}{\langle}
\newcommand{\ra}{\rangle}
\newcommand{\amp}[1]{\la #1 \ra}
\newcommand{\Tr}{{\text{Tr}}}

\newcommand{\slsh}[1]{\mbox{$\not\! #1$}}
\newcommand{\bm}[1]{\bbox{#1}}

\newcommand{\dbrl}{\Big[\!\!\Big[}
\newcommand{\dbrr}{\Big]\!\!\Big]}

\begin{document}
 
\draft
\title{
\begin{flushright}
\begin{minipage}{4 cm}
\small
hep-ph/9906223\\
RIKEN-BNL preprint\\
VUTH 99-12
\end{minipage}
\end{flushright}
Color gauge invariance in the Drell-Yan process}

\author{Dani\"el Boer$^1$ and P.J.\ Mulders$^2$}
\address{\mbox{}\\
$^1$RIKEN-BNL Research Center\\
Brookhaven National Laboratory, Upton, NY 11973, U.S.A.\\
\mbox{}\\
$^2$Department of Physics and Astronomy, Free University \\
De Boelelaan 1081, NL-1081 HV Amsterdam, the Netherlands
}

\maketitle
\begin{center}\today \end{center}

\begin{abstract}
We consider the color gauge invariance of a factorized description of the 
Drell-Yan process cross section. In particular, we focus on the 
next-to-leading twist contributions for polarized scattering 
and on the cross section differential in 
the transverse momentum of the lepton pair in the region where the transverse
momentum is small compared to the hard scale. 
The hadron tensor is expressed in 
terms of manifestly color gauge invariant, nonlocal operator matrix elements 
and a color gauge invariant treatment of soft gluon poles is given. Also, we 
clarify the discrepancy between two published results for 
a single spin asymmetry in the Drell-Yan cross section. This asymmetry 
arises if such a soft gluon pole is present in a specific twist-three
hadronic matrix element.
\end{abstract}


\section{Introduction}

\noindent 
In this article we will address a number of issues concerning the use
of nonlocal matrix elements to describe the nonperturbative `soft' parts 
that enter in the calculation of hard scattering cross sections. 
Specifically we consider color gauge invariance of these matrix elements in
case effects of intrinsic transverse momentum of partons are included, 
necessarily involving nonlocality away from the lightcone. 
We will study in detail the next-to-leading order in an expansion in inverse 
powers of the hard scale $Q$ for polarized scattering, i.e., order $1/Q$, 
also referred to as twist-three. This is for instance relevant in the study of
the single spin asymmetry investigated in Refs.\ \cite{Hammon-97} and 
\cite{Boer4}.

In a factorized description of hadron-hadron scattering processes like the
Drell-Yan (DY) process, one can work with matrix elements containing nonlocal
operators (the local operator product expansion does not apply). 
At the leading order in inverse powers of the hard scale, 
operators containing arbitrary numbers of gluons with polarization collinear to
the momentum of the parent hadron 
need to be combined in order to render such matrix elements with nonlocal
operators color gauge 
invariant. The gluon contributions sum up to produce
path-ordered exponentials, also called link operators, with paths running 
along finite distances in between the fields that are situated along a 
lightlike direction. 

Application of Ward identities to gluons with polarization along their
momenta, so-called longitudinally polarized gluons, is an important 
ingredient to arrive at these path-ordered exponentials. At the leading twist,
but  
also beyond leading order in inverse powers of the hard scale, one 
can often perform a so-called collinear expansion of the hard scattering part, 
such that the partons that
need to be considered only have momenta collinear to the parent hadron. 
In that 
case longitudinally polarized gluons are polarized collinear to 
the hadron momentum, allowing for straightforward application of Ward
identities.  However, in some cases (e.g.\ azimuthal asymmetries) one has to 
deal with partonic transverse momenta and the resulting difference between 
collinear and longitudinal polarization of gluons 
starts to complicate matters. This difference involves matrix elements
containing 
transverse gluon fields and therefore, is only relevant beyond leading order 
in inverse powers of the hard scale. Such a case is for instance the DY
cross section differential in 
the transverse momentum of the lepton pair in the region where the transverse
momentum is small compared to the hard scale. Despite the mentioned 
complication, we will show for this case that appropriate 
path-ordered exponentials will result nevertheless, i.e.\ also at the
next-to-leading twist (${\cal O}(1/Q)$). 

In case the
hadronic matrix elements have a dependence on the transverse momenta of the
partons, the nonlocality of the operators is forced to be off the light-cone. 
Consequently, the path-ordered exponentials that are needed to render these 
nonlocal operator matrix elements color gauge invariant, involve paths that 
are also off the light-cone. However, to the order considered here, one finds 
(as will be shown below) several lightlike paths which are transversely 
separated from each other, depending on the transverse position of the
colored fields they connect to. The paths 
extend to lightcone infinity. Such paths are not uncommon, e.g.\ 
\cite{CSS83,Col-89,Balitsky-Braun-91}, but it has not been demonstrated
before that they also arise in the case(s) under consideration here.

Hadronic matrix elements involving fields with their 
arguments at lightcone infinity are usually assumed to vanish. In the case of 
so-called gluonic or soft gluon pole contributions, which are related to 
transverse gluon 
fields at infinity \cite{Boer4}, this issue needs careful reconsideration. We
find that a color gauge invariant treatment of soft gluon poles in  
twist-three hadronic matrix elements
\cite{QS-91b,Korotkiyan-T-94,E-Korotkiyan-T-95} can be given
also.  

In addition to these general results on color gauge invariant matrix elements
of nonlocal operators in the description of the 
Drell-Yan cross section differential in the transverse
momentum of the lepton pair, we will clarify the 
apparent discrepancy between the
results for the power-suppressed single transverse spin azimuthal asymmetry 
in the DY cross section as discussed in Refs.\ \cite{Hammon-97} and 
\cite{Boer4}. 
In Ref.\ \cite{Hammon-97} this specific single spin asymmetry is assumed to
arise from such a soft gluon pole contribution. In Ref.\ \cite{Boer4} it is
shown that the same asymmetry can also arise from a so-called time-reversal 
odd distribution function. It was shown that although time reversal 
symmetry prohibits the appearance of such functions (if one considers the
incoming hadron states to be plane wave states) the soft gluon pole matrix 
element can be treated as an effective time-reversal odd distribution 
function, without violating time reversal symmetry. Nevertheless, there
remained a discrepancy between the results in Refs.\ \cite{Hammon-97} and 
\cite{Boer4} arising from an assumption on the factorized 
expression for the hadron tensor. We will make this 
assumption explicit and carefully study this issue. 
First we will discuss the discrepancy between the two single spin asymmetry 
expressions in a bit more detail. 

\section{Single spin asymmetry in the Drell-Yan process}

The final result for the single spin asymmetry in the DY cross section
arising from a soft gluon pole contribution in a particular 
twist-three hadronic matrix element
\cite{QS-91b,Korotkiyan-T-94,E-Korotkiyan-T-95} as presented in Ref.\ 
\cite{Hammon-97} is of the form (Eq.\ (15) in \cite{Hammon-97}):
\beq
A_1= g \, \frac{\sin 2\theta \cos \phi \left[ T(x,x) -x \, 
\frac{dT(x,x)}{dx}\right]
}{Q \left[1+ \cos^2 \theta\right]q(x)}.
\label{HammonA}
\eeq
The unpolarized quark distribution $q(x)$ and the twist-three soft gluon pole
function $T(x,x)$ depend on the lightcone momentum fraction $x$ of the quark
inside the transversely polarized hadron only. 
The function $T(x,x)$ is up to a minus sign identical to
the function $T(x,S_T)$ introduced in Ref.\ \cite{QS-91b}. We have
chosen to denote the mass of the lepton pair by $Q$, instead of $M$ as was
done in \cite{Hammon-97}, in
order to avoid confusion with a hadron mass. The angles are defined in the
dilepton center of mass frame, cf.\ Ref.\ \cite{Hammon-97}.
 
In Ref.\ \cite{Boer4} a similar result was derived  
from another perspective, namely that of so-called time-reversal odd 
distribution functions. 
The asymmetry (Eq.\ (71) in \cite{Boer4}) was found to be\footnote{We have 
corrected the proportionality constant, i.e.\ we have replaced $4$ by $-2$.}  
(just as Eq.\ (\ref{HammonA}) given in the dilepton center of mass frame): 
\beq
A_2 = \frac{-2\, \sin 2\theta \,\sin \phi_{S_1}}{1 + \cos^2\theta} 
\frac{|\bm S_{1T}^{}|}{Q} \frac{\sum_{a}e_a^2
\;\bigg[M_1 \, x_1 \,f_T^a (x_1)\, 
f_1^{\bar a}(x_2) + M_2\, x_2\, h_1^a(x_1) \, 
h^{\bar a}(x_2) 
\bigg]}{\sum_{a}e_a^2 \; f_1^a (x_1) f_1^{\bar a} (x_2)}. 
\label{BoerA}
\eeq
The first term in the asymmetry (proportional to $f_T$) equals the term 
proportional to $T(x,x)$ in Eq.\ (\ref{HammonA}).
To be more explicit, 
\ba
&& x {f}_T(x) = 
\frac{g}{2 M S_T^2} T(x,x)=\frac{-g \, \pi}{2 M S_T^2 P^+ } 
\text{Tr}\left[ \Phi_F^\alpha (x,x)\, \slsh{n_-}
\right] \epsilon_{T}^{S_T \alpha },
\label{defT}\\[2 mm]
&& x {h}(x) = \frac{- i \pi \, g}{2MP^+}
\text{Tr}\left[ \Phi_F^\alpha (x,x)\, \gamma_{T \alpha} \slsh{n_-}\right], 
\ea 
where $\epsilon_{T}^{S_T \alpha}=\epsilon^{S_T \alpha n_+ n_-}=
\epsilon^{\beta \alpha \rho\sigma} S_{T\beta} n_{+\rho} n_{-\sigma}$ and 
the correlation function $\Phi_F^\alpha$ {\em in the $A^+=0$ gauge\/} 
is given by
\beq
\Phi_{Fij}^\alpha(x,y) \equiv \int \frac{d \lambda}{2\pi} 
\frac{d \eta}{2\pi} e^{i\lambda x} e^{i\eta (y-x)} 
\amp{P,S|\psibar_j (0) F^{+\alpha} (\eta n_-) 
\psi_i(\lambda n_-)| P,S}.
\label{defPhiF}
\eeq
The hadron momentum is up to a mass term proportional to $n_+$, which is 
one of two lightlike vectors $n_+$ and $n_-$, chosen such that 
$n_+\cdot n_-=1$. We will often refer to the $\pm$ components of a vector $p$,
which are defined as $p^\pm=p\cdot n_\mp$. To be more specific, we have 
decomposed the hadron momentum and spin vectors $P$ and $S$ as
\begin{eqnarray}   
&& P^\mu \equiv \frac{ Q}{x \sqrt{2}}\,n_+^\mu 
+ \frac{x M^2}{ Q\sqrt{2}}\,n_-^\mu,\\
&& S^\mu  \equiv \frac{\lambda Q}{x M \sqrt{2}}\,n_+^\mu
-\frac{x \lambda M}{Q \sqrt{2}}\,n_-^\mu + S_{T}^{\mu}.
\end{eqnarray}

The second term in Eq.\ (\ref{BoerA}) is another soft gluon pole 
contribution to the same single spin asymmetry in the DY process. It
is not proportional to $T(x,x)$, but to a chiral-odd projection of
$\Phi_F^\alpha$ at the point $x=y$. Apart from this chiral-odd term, 
the difference between $A_1$ and $A_2$ is in the additional term in $A_1$
proportional to $x \, dT(x,x)/dx$. 
To see the discrepancy more clearly we will look at Eq.\ (\ref{BoerA}) in 
case of one
flavor, a purely transversely polarized hadron, i.e., $|\bm S_{1T}^{}|=1$, 
and disregard the chiral-odd term. The definitions of the azimuthal angles
$\phi$ and $\phi_{S_1}$ differ by $\pi/2$ and the unpolarized quark 
distribution $q(x) = f_1(x)$. Hence, one has for the asymmetry 
\beq
A_2 = g \, \frac{\sin 2\theta \cos \phi \, T(x,x)
}{Q \left[1+ \cos^2 \theta\right]q(x)},
\label{BoerA2}
\eeq
which indeed differs from $A_1$ given in Eq.\ (\ref{HammonA}) by the term 
proportional to $x \, dT(x,x)/dx$. Such a derivative term was indeed found  
in other processes, like prompt photon production \cite{QS-91b} and 
pion production in proton-proton scattering \cite{QS-98}, where it arises 
from collinear expansions of the hard scattering part. 
As we will show in the next section, the derivative term obtained in
Ref.\ \cite{Hammon-97} for  
the particular case of the DY process {\em at the 
tree level\/}, arises in a different way, namely from the 
unnecessary 
requirement of $n_-$-independence of the Fierz decomposed hadron tensor. 
Of course, such a derivative term {\em can\/} arise 
from collinear expansions of the hard scattering parts 
{\em beyond tree level\/}. 

Before we discuss this problem, we will first 
carefully
consider the color gauge invariance of the description of the cross section,
which is an important issue in itself.
We will first review the case of 
deep inelastic scattering (DIS), after
which we return to the analysis of the DY process. 

\section{\label{II}Deep Inelastic Scattering}

\subsection{\label{IIA}Color gauge invariance}

\noindent 
We will consider the description of the deep inelastic scattering 
cross section in terms of color gauge invariant functions involving nonlocal
operators. The hadron tensor $W^{\mu\nu}$ can be written as 
\cite{EFP-83,Efremov-Teryaev-84} 
\beq
W^{\mu\nu}= \int dx \, \Tr \left[ E^{\mu\nu} (x) \Phi (x) \right] + 
\int dx_1 dx_2 \, 
\Tr \left[ E^{\mu\nu}_\alpha (x_1,x_2) {\omega^{\alpha}}_{\beta} 
\Phi_D^\beta (x_1,x_2) \right]
+ \cdots
\label{efp1}
\eeq
Here, $E^{\mu\nu}, E^{\mu\nu}_\alpha$ and $\Phi,\Phi_D^\beta$ are hard and 
soft scattering parts, respectively. The hard part $E^{\mu\nu}$ is the
discontinuity of the 
$\gamma^*$--quark forward scattering amplitude and $E^{\mu\nu}_\alpha$ has an 
additional gluon connected to it. They are 1PI in the 
external legs. The color gauge invariant soft parts are defined as
\ba
\Phi_{ij} (x) &\equiv& \int \frac{d \lambda}{2\pi}e^{i\lambda x}\amp{P,S|\,
\psibar_j (0) \, {\cal L}^+ [0,\lambda n_-] \, \psi_i(\lambda n_-) | P,S}
\label{PhiDIS}\\[2 mm]
\Phi_{Dij}^{\alpha} (x_1,x_2) &\equiv&
\int \frac{d \lambda}{2\pi} 
\frac{d \eta}{2\pi} e^{i\lambda x_1} e^{i\eta (x_2 -x_1)} \amp{P,S|\,
\psibar_j (0) \, {\cal L}^+ [0,\eta n_-] \, iD^{\alpha}(\eta n_-) 
\, {\cal L}^+ [\eta n_-,\lambda n_-]\, \psi_i(\lambda n_-) |
P,S},\label{PhiDDIS}
\ea
where 
\beq
{\cal L}^+ [\eta n_-,\lambda n_-] = {\cal P} 
\exp \left(-ig\int_{\eta}^{\lambda} d \zeta \, \left(n_-\cdot A(\zeta n_-) 
\right) \right).
\label{linkDIS}
\eeq
The projector ${\omega^{\alpha}}_{\beta}={g^{\alpha}}_{\beta}-n_+^{\alpha} 
n_{-\beta}$ implies that the index of the covariant derivative is either 
transverse or in the minus-direction (the latter contributes at order 
$1/Q^2$ and will be neglected here).   
The above result holds for DIS up to order $1/Q$ and is relevant for 
polarized scattering. 
It was arrived at from a collinear expansion of the starting 
expression\cite{EFP-83,Efremov-Teryaev-84} 
\beq
W^{\mu\nu} = \int d^4k \, \Tr \left[ E^{\mu\nu} (k) \Phi (k) \right] 
+ \int d^4 k_1 \, d^4k_2 \, \Tr \left[ E^{\mu\nu}_\alpha 
(k_1,k_2) \Phi_A^\alpha (k_1,k_2) \right] + \ldots,
\label{Tmunu}
\eeq
without the projector $\omega$ and
with color gauge {\em variant\/} Green's functions  
\ba
\Phi_{ij}(k)&=&\dz e^{ik \cdot z} \amp{P,S|T \, \psibar_j (0) \psi_i (z) 
 | P,S},\\[2 mm] 
\Phi^{\alpha}_{Aij}(k_1,k_2)&=&\dz \dzp e^{ik_1 \cdot z} e^{i(k_2-k_1) \cdot 
z'} \amp{P,S|T \, \psibar_j (0) g A^{\alpha}(z') \psi_i (z) | P,S}.
\ea
The collinear direction $n_+$ is defined by the hadron momentum (up to a
hadron mass term) and the parton
momenta are decomposed as 
\beq
k_i \equiv \frac{x_i Q}{\sqrt{2}}\,n_+
+ \frac{(k_i^2 + \bm{k}_{iT}^2)}{x_i Q \sqrt{2}}\,n_- + k_{iT}.
\eeq
From power counting one sees that the plus-component of $\Phi^{\alpha}_{A}$
can start to contribute at order 1. The transverse and minus components are
order $1/Q$ and $1/Q^2$ respectively. In a general gauge one writes 
$A^\alpha=n_+^\alpha \,
(n_-\cdot A) + 
{\omega^\alpha}_\beta \, A^\beta$, such that the term proportional to
$n_+^\alpha$ will yield
\ba
\lefteqn{\int d x_1 d x_2 \, \Tr \left[ E^{\mu\nu}_\alpha(x_1, x_2) \, 
n_+^\alpha \,  
n_{-\beta} \Phi_{A}^\beta (x_1,x_2) \right] =}\nn\\
&& \qquad \int dx\, \Tr \left[E^{\mu\nu}(x) 
\int \frac{d\lambda}{2\pi} e^{i\lambda \, x} \amp{P| \, \psibar(0) 
\left(-ig\int_0^\lambda \, d\zeta\, \left( n_-\cdot A(\zeta n_-) 
\right) \right) 
\psi(\lambda n_-) |P} \right],
\ea
due to the Ward identity \cite{EFP-83}
\beq 
\frac{Q}{\sqrt{2}} \, n_+^\alpha \, E^{\mu\nu}_\alpha (x_1,x_2) =
\frac{E^{\mu\nu}(x_2)-E^{\mu\nu}(x_1)}{x_2-x_1}.
\label{WardDIS}
\eeq
Combined with matrix elements containing arbitrary numbers of 
$A^+$ gluons (all
contributing at the same order, which in this case is order 1), this
will exponentiate into the path-ordered exponential in $\Phi(x)$
\cite{Rad-Efr,Coll-S-82,CSS89,Qiu-Sterman-91}, with a
straight path along the $n_-$-direction, Eq.\ (\ref{linkDIS}).
Hence, color gauge invariance of the
factorized hadron tensor is manifest and for instance, 
a lightcone gauge $A^+=0$ can be
chosen without problems, reducing the path-ordered exponential to unity. 

We note that one starts out with the correlation function
$\Phi_A^\alpha$, without restriction on $\alpha$, 
and then derives an expression
involving $\Phi_D^\alpha$, where only the transverse components are 
contributing up to order $1/Q^2$ corrections.  
The parametrization of $\Phi_D^\alpha(x_1,x_2)$ can then be chosen to 
reflect the fact that only the transverse part is contributing. 
This is the course taken in Ref.\ 
\cite{Jaffe-Ji-91}, which uses the parametrization
\begin{eqnarray}
\Phi^{\alpha}_{D}(x,y)&= & 
\frac{M}{2} \bigg[ G_D(x,y)\,i\epsilon_T^{\alpha\beta} 
S_{T \, \beta} \slsh{P} + \tilde{G}_D(x,y)\,
S_{T}^\alpha \gamma_{5} \slsh{P} + H_D(x,y) \lambda\gamma_{5} 
\gamma_T^\alpha \slsh{P} 
+E_D(x,y)  \gamma_T^\alpha \slsh{P} \bigg],
\label{paramPhiD}
\ea
where $\epsilon_T^{\mu\nu}=\epsilon^{\alpha \beta\mu\nu} n_{+\alpha}
n_{-\beta}$. 
Obviously, this parametrization depends on $n_-$, nevertheless, one 
will obtain an $n_-$-independent cross section if $n_-$-independence (or
equivalently Lorentz invariance) is imposed on the functions in the starting
expression Eq.\ (\ref{Tmunu}). We will address this issue of
$n_-$-independence and its consequences in the next section. 

\subsection{Lorentz invariance and $n_-$-independence}

We have shown that choosing a lightcone gauge $n_- \cdot A =0$ is not 
required, 
hence, the vector $n_-$ is not (necessarily) associated with the choice of 
gauge. The
vector $n_-$ reflects the choice of basis in which the hard scattering 
process is expressed. The requirement of $n_-$-independence of the {\em
final\/} result
is then simply the requirement of Lorentz invariance. Since the Lorentz
covariant tensor $W^{\mu\nu}$ is written in terms of traces of 
hard and soft parts {\em after\/}
the choice of this basis, $n_-$-independence of the
hard and soft parts separately is {\em not\/} a requirement. 
For a different choice of
$n_-$, one would obtain a different expression for the hard and soft parts and
only the trace of the two parts will be $n_-$ independent (which we have 
explicitly confirmed for the calculation of the DY asymmetry in 
Ref.\ \cite{Boer4}). The reason that the
choice of $n_-$ has to be made is because in DIS there are only two external
momenta, namely $P$ and $q$, hence not all directions that arise in the
calculation can be expressed 
solely in terms of external momenta. In the DY process this {\em can\/} be done
using $P_1, P_2, q$ and $(P_1\times P_2) \times q$, hence one need not resort
to an arbitrary choice of $n_-$ in principle. Therefore, in the case of the DY
process the issue of $n_-$-independence can be avoided altogether.  

The expression for the hadron tensor in Ref.\ \cite{Hammon-97} (taken from
Ref.\ \cite{Efremov-Teryaev-84}) differs from Eq.\ (\ref{efp1}) (after 
insertion
of the parametrizations of $\Phi$ and $\Phi_D^\alpha$), due to the additional 
requirement of $n_-$-independence of the factorized hadron
tensor expression Eq.\ (\ref{efp1}). 
To be explicit, the expression for the hadron tensor in Ref.\ 
\cite{Hammon-97} (their Eq.\ (1)) is given by (we have replaced the index
$\mu$ by $\alpha$ and the hadron momentum $p$ by $P$ to avoid confusion)
\ba
W&=& \int dx\, C_T \frac{1}{4} \Tr \left[ E(x) \slsh{S} \gamma^5
\right]\nn\\[2 mm]
&+& \int dx_1\, dx_2 \, B^A(x_1,x_2) \frac{1}{4} \Tr \left[ E_\alpha(x_1,x_2)
\slsh{P} \gamma^5 \right] S^\alpha +
\int dx_1\, dx_2 \, B^V(x_1,x_2) \frac{1}{4} \Tr \left[ E_\alpha(x_1,x_2)
\gamma_\rho \right]\epsilon^{\rho\alpha S P}, 
\label{HammonHadron}
\ea 
with $\epsilon^{\rho\alpha S P}= \epsilon^{\rho \alpha \mu \nu} S_\mu
P_\nu$.
The functions 
$E$ and $E_\alpha$ are the hard scattering parts $E^{\mu\nu}$ and
$E_\alpha^{\mu\nu}$ of Eq.\ (\ref{efp1}). The Dirac 
projections of the soft parts $\Phi(x)$ and $\Phi_D^\alpha(x,y)$ are 
denoted by the functions $C_T, B^A$ and $B^V$ and hence constitute the
parametrization of $\Phi_D^\alpha$. The function $B^V$ is related 
to the function $G_D$ of Eq.\ (\ref{paramPhiD}) as (cf.\ Eq.\ (4) of
\cite{Hammon-97}) 
\beq
B^V(x_1,x_2)=-i \Tr\left[ \Phi_{D\,\alpha} (x_2,x_1) \slsh{n_-} \right]
\epsilon^{\alpha S n_+ n_-}/P^+ = 2 M S_T^2 \, G_D(x_2,x_1).
\eeq 
The function $B^A$ is related to the function $\tilde{G}_D$ and 
the chiral-odd functions $H_D$ and $E_D$ are not considered in 
Eq.\ (\ref{HammonHadron}).

The last term in Eq.\ (\ref{HammonHadron}) 
is the one that is relevant for our discussion, since the soft gluon poles 
reside in the function $B^V$. The relation to 
$T(x,x)$ is given by 
\beq
T(x,x)=\lim_{y\rightarrow x} T(x,y) = \lim_{y\rightarrow x} \left[ 
\pi B^V(x,y) \, (x-y) \right]/g,
\eeq  
which shows that if $T(x,x)$ is nonzero, the function $B^V(x,y)$ has to have a
pole at the point $x=y$. 

The last term in Eq.\ (\ref{HammonHadron}) is accompanied by the term 
$\epsilon^{\rho\alpha S P} \sim \epsilon^{\rho\alpha S n_+}$. 
Next we note the identity 
\beq
\epsilon^{\rho\alpha S n_+}= n_+^\alpha \, 
\epsilon^{\rho S_T n_+ n_-} - n_+^\rho \, \epsilon^{\alpha S_T n_+ n_-}
-(S\cdot n_+)\, \epsilon^{\rho\alpha n_+ n_-}. 
\label{Schouten}
\eeq
The second and third term on the r.h.s.\ restrict the index $\alpha$ 
to be transverse, unlike the first term, which 
implies that $\alpha$ can be $+$. The first term potentially produces a 
contribution in the hadron tensor Eq.\ (\ref{HammonHadron}) 
that is proportional to $B^V$ and 
hence $G_D$, whereas the projector $\omega$ in the hadron 
tensor Eq.\ (\ref{efp1}) prohibits such a contribution. 

The expression for the hadron tensor 
Eq.\ (\ref{HammonHadron}) can be written in the form of Eq.\ (\ref{efp1}) 
without $\omega$ projector, in which case
$\Phi_D$ is given by
\beq
\Phi^{\alpha}_{D}(x,y)= 
\frac{1}{4} \bigg[ B^V(x,y)\,\epsilon^{\rho \alpha S P} \, \gamma_\rho 
+ B^A(x,y)\, S^\alpha \slsh{P} \gamma_{5} \bigg],
\eeq
which is equivalent to imposing $n_-$-independence 
on the parametrization of $\Phi_D^\alpha$ ({\em without\/} the $\omega$
projector), which is clearly different from  
Eq.\ (\ref{paramPhiD}) (disregarding the chiral odd pieces). 
As said before this will potentially imply the appearance of 
additional contributions.
In DIS such additional contributions happen 
not to arise\footnote{This is in fact due to 
time reversal invariance. 
The derivation of Eq.\ (\ref{HammonHadron}) is 
given in Ref.\ \cite{Efremov-Teryaev-84} and requires $n_-$ independence
of the Fierz decomposed hadron tensor. 
The $n_-$-dependent terms are required to cancel each
other, but the resulting equation (Eq.\ (43b) in Ref.\
\cite{Efremov-Teryaev-84}) involving the function $B^V$ turns out to 
vanish due to time reversal invariance.}, 
but in the DY process it will be the case. So the two expressions for the hadron tensor (Eq.\ (\ref{efp1}) and
Eq.\ (\ref{HammonHadron})) give the same
results in DIS, but they do yield different results in the DY process
calculation. In fact, the first 
term in Eq.\ (\ref{Schouten}) is the origin of the $x \, dT(x,x)/dx$ term 
in the asymmetry expression $A_1$ Eq.\ (\ref{HammonA}).   

In conclusion, the requirement of $n_-$-independence 
of the hadron tensor Eq.\ (\ref{efp1}) will lead to a different answer for 
the Drell-Yan cross section in the case soft gluon poles are present. 
Without imposing this unnecessary requirement the derivative term will not 
be present in the asymmetry expression. 
Let us emphasize that this does not apply to the
derivative terms obtained from collinear expansions of the hard parts, such as
arising in prompt photon production \cite{QS-91b},  
pion production in proton-proton scattering \cite{QS-98} and (presumably) 
the DY process {\em beyond\/} tree level.    

\section{Color Gauge Invariance in the Drell-Yan process}

\noindent 
In section \ref{IIA} on DIS we discussed how the collinear polarization  
part of any number of gluons (the $A^+$ gluons) exponentiate using a Ward
identity rendering the nonlocal operators color gauge invariant. 
In analogy to the starting point, Eq.\ (\ref{Tmunu}), for DIS, 
for the Drell-Yan process the starting expression for the hadron tensor 
integrated over the transverse momentum $\bm q_T$ of the
lepton pair and including order $1/Q$ contributions, is given by 
\begin{eqnarray}
&&\int d^2 \bqt \, {\cal W}^{\mu\nu}(x,\bar{x}) 
= \frac{1}{3} \,\Biggl\{ 
\text{Tr}\left( \Phi (x) \gamma^\mu \overline \Phi 
(\bar{x}) \gamma^\nu \right) \label{qtWmunu}\\
&& \quad
+ \int\, dy \, \text{Tr}\left( \Phi_A^\alpha (y,x) \gamma^\mu \overline \Phi 
(\bar{x}) \gamma_\alpha \frac{\slsh{n_+}}{Q\sqrt{2}} \, \frac{x-y}{x-y + i
\epsilon}\, \gamma^\nu \right) +
\int\, dy \, \text{Tr}\left( \Phi_A^\alpha (x,y)  \gamma^\mu
 \frac{\slsh{n_+}}{Q\sqrt{2}} \, \frac{x-y}{x-y - i\epsilon}\,
\gamma_\alpha \overline \Phi (\bar{x}) \gamma^\nu 
\right)
\nonumber \\[3mm]
&& \quad -\int\, d\bar{y} \, \text{Tr} \left( \Phi (x) \gamma^\mu  
\overline \Phi_A^\alpha (\bar{y},\bar{x}) \gamma^\nu 
\frac{\slsh{n_-}}{Q\sqrt{2}} \, \frac{\bar{x} -\bar{y}}{\bar{x} -\bar{y}+ 
i \epsilon}\, \gamma_\alpha 
\right) - \int\, d\bar{y} \, \text{Tr}\left( \Phi(x) \gamma_\alpha
 \frac{\slsh{n_-}}{Q\sqrt{2}} \, \frac{\bar{x} -\bar{y}}{\bar{x} -\bar{y}- 
i \epsilon}\, \gamma^\mu \overline \Phi_A^\alpha (\bar{x}, 
\bar{y}) \gamma^\nu 
\right) \Biggr\}.\nn 
\end{eqnarray}
In this expression the terms with $\slsh{n_\pm}$ arise from the fermion 
propagators in the hard part neglecting contributions that will appear 
suppressed by $1/Q^2$ (for details see Ref.\
\cite{Boer4}\footnote{Note that two 
misprinted signs for the $i \epsilon$ pole prescriptions as given in 
Ref.\ \cite{Boer4} have been corrected, the 
results there were produced with the signs given here.}). In case one assumes
that the functions $\Phi_A^\alpha (x,y)$ (and 
$\overline \Phi_A^\alpha (\bar{x}, \bar{y})$) to be regular in the point $x=y$,
where the gluon has zero momentum, one can replace terms like  
\beq
\frac{\slsh{n_+}}{\sqrt{2} Q} \frac{x-y}{x-y+i\epsilon}
\to \frac{\slsh{n_+}}{\sqrt{2} Q}. 
\eeq
In sections \ref{IIIA} and \ref{IIIA2} 
we will assume that this is indeed the case and we will
return to this issue in section \ref{IIIB}, when we investigate the
consequences of poles in $\Phi_A^\alpha (y,x)$ at the point $x=y$, the
so-called soft gluon poles. 

Below we explain how Eq.\ (\ref{qtWmunu}) can be expressed in terms of 
manifest color gauge invariant matrix elements by summing collinear 
gluons. One conclusion will be that, also with the effects of transverse
momenta included, color gauge
invariance implies that the index on $\Phi_D^\alpha$ is in fact transverse.

The issue of color gauge invariance in processes with two soft parts, like the
DY process, is a theoretical topic not yet fully addressed in the 
literature. One
has to show that a process with two soft parts factorizes into color gauge
invariant objects. We note that if the cross section factorizes into 
color gauge invariant functions, this will allow one to use different 
gauges for different correlation functions in the same diagram.

For DIS, a process with one soft part, Ward identities applied to
correlation functions with arbitrary numbers of $A^+$-gluons, yield the 
desired path-ordered exponentials or link operators, that
render the correlation functions color gauge invariant (see the previous 
section). 
In \cite{Rad-Efr,CSS85-b,Qiu-Sterman-91} this issue was considered for the DY 
process, but the transverse
momentum of the quarks is not included in the most general way. Unlike
\cite{Rad-Efr,CSS85-b,Qiu-Sterman-91} we will focus on the case that the 
transverse momentum of the lepton pair is not integrated over and is 
small compared to the hard scale. 
In Ref.\ \cite{CSS83,CSS85} 
transverse momentum {\em is\/} taken into
account, but only for the leading twist. 
We will first 
consider tree level in the
hard scattering part, but all orders in the coupling constant 
appearing in the link
operators. Next we will discuss the extension of these results to 
all orders in $\alpha_s$ in the hard scattering part by means of 
Ward identities. Our analysis will show explicitly how
collinear Ward identities can be applied to the case of the DY process with 
transverse momentum, including next-to-leading twist, in order to show that
collinear gluons exponentiate. This will also be an important ingredient in a
future proof of exponentiation of soft gluons into Sudakov factors. 
  
\subsection{\label{IIIA}Color gauge invariant correlation functions at tree
level}

We will now show that the DY process including 
$1/Q$ contributions, actually consists of color gauge invariant correlation
functions containing link operators with straight paths along a lightlike
direction extending to lightcone infinity. Such a link operator in
$\Phi(p)$ is for each colored field of the form 
\begin{equation}
{\cal L}^+[-\infty,x] = \left. {\cal P} \exp \left(- ig \int_{-\infty}^{x^-} 
dy^- \, A^+(y) \right)\right|_{y^+ = x^+, \bm{y}_T=\bm{x}_T}.
\label{linkDY}
\end{equation}
We will extensively study 
the tree level situation and note that the coupling constants appearing in the
link operators are part of the color gauge invariant definition of the 
correlation
functions, hence, count as tree level objects and not as part of the
perturbative corrections to the hard scattering. 
We will consider the case of the DY process, but the case of two-hadron 
production in electron-positron annihilation or semi-inclusive DIS 
will be completely analogous, although the direction in which the links are
running is different. 
For instance, in the case of leptoproduction one finds for each field
in the correlation function $\Phi(p)$ the link operators
\begin{equation}
{\cal L}^+[\infty,x] = \left.
{\cal P} \exp \left(ig \int_{x^-}^\infty dy^- \, A^+(y)
\right)\right|_{y^+ = x^+, \bm{y}_T = \bm{x}_T}.
\end{equation}

Since the cross section as a whole is gauge invariant, we will
choose {\em one\/} lightcone gauge $A^-=0$ and show that 
summation of $A^+$ gluons
leads to appropriate link operators in one set of correlation functions (the
correlation functions $\Phi, \Phi_D^\alpha$). This results in the color gauge
invariance of these correlation functions and thereby we will have  
demonstrated that choosing a lightcone gauge in correlation functions is 
not necessary (any or no gauge choice will do). 

In the $A^-=0$
gauge one has to include $A^+$ gluons in all possible ways. 
We observe that 
$A^+$ gluons emanating from $\overline \Phi$ will give rise to $1/Q^2$ 
suppression, so we only need to consider the cases where the $A^+$ gluons 
emanate from $\Phi$. 
We will show that matrix elements with multiple $A^+$-gluon fields 
in $\Phi^{\,+}_{\,A}, \Phi^{\,++}_{\,AA}, \ldots$ and 
$\Phi^{\,\alpha +}_{\,AA}, \Phi^{\,\alpha ++}_{\,AAA},
\ldots$
will combine into link operators in $\Phi$ and $\Phi^\alpha_D$, 
with paths as given above (running along
the minus direction extending to $y^-=-\infty$). 
Note that the correlation function 
$\Phi^\alpha_D$ appears in the end, since the explicit $A^\alpha$ 
in $\Phi_A^\alpha$ is not gauge invariant. We also note that in the $A^+=0$ 
gauge one has (due to $iD^{\alpha}= i\partial^{\alpha} + g
A^{\alpha}$) 
\begin{equation}
\Phi_{A}^{\alpha} (p) \equiv \int d^4 p_1\, \Phi_{A}^{\alpha} (p_1,p) = 
\int d^4 p_1\, \Phi_{D}^{\alpha} (p_1,p) - p^\alpha \,\Phi(p),
\label{p1integratedPhi}
\end{equation}
but for color gauge invariant correlation functions this relation will be
modified, since $i\partial^\alpha$ also acts on link operators. 

We will first consider the case where all gluons
connect to the left side of the diagram. 
In Fig.\ \ref{dyfig} we depict one such diagram that is
relevant for the tree level
calculation including $1/Q$ corrections \cite{Boer4}. 
\begin{figure}[htb]
\begin{center}
\leavevmode \epsfxsize=6cm \epsfbox{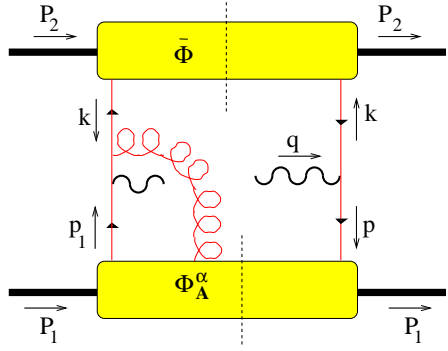}
\caption{\label{dyfig}A diagram contributing to the Drell-Yan
process at order $1/Q$.}
\end{center}
\end{figure}
The diagrams with first of all {\em zero or one\/}  
$A^+$ gluon(s) that we need to consider are (schematically) 
displayed in Fig.\ \ref{links}. We have only drawn the relevant 
left part of the diagrams and left out the blobs denoting the correlation
functions, so gluons going to the bottom (top) emanate from 
$\Phi$ ($\overline \Phi$). 
For instance, Fig.\ \ref{links}b represents Fig.\ \ref{dyfig}.
The index $\alpha$ in Fig.\ \ref{links}b can be 
either transverse ($T$) or $+$ and it turns out
that $\alpha$ and $\beta$ in Fig.\ \ref{links}c will be $T$ and 
$+$ respectively,
otherwise the diagram will give a suppressed contribution; 
on the other diagrams we 
have indicated the appropriate gluon polarization (collinear or transverse). 

The diagrams (h) and (i) are the only diagrams with a triple gluon vertex,
since others will either be of higher order in $\alpha_s$ or cannot
be separated from the soft part, since not all momenta are participating in 
the hard scattering and soft momentum loops will appear. We will first ignore
the issue of color matrices in the diagrams without a triple gluon vertex and
return to this point after we have considered the diagrams (a)-(g), i.e.,
after Eq.\ (\ref{ef}).  

\begin{figure}[htb]
\begin{center}
\leavevmode \epsfxsize=11cm \epsfbox{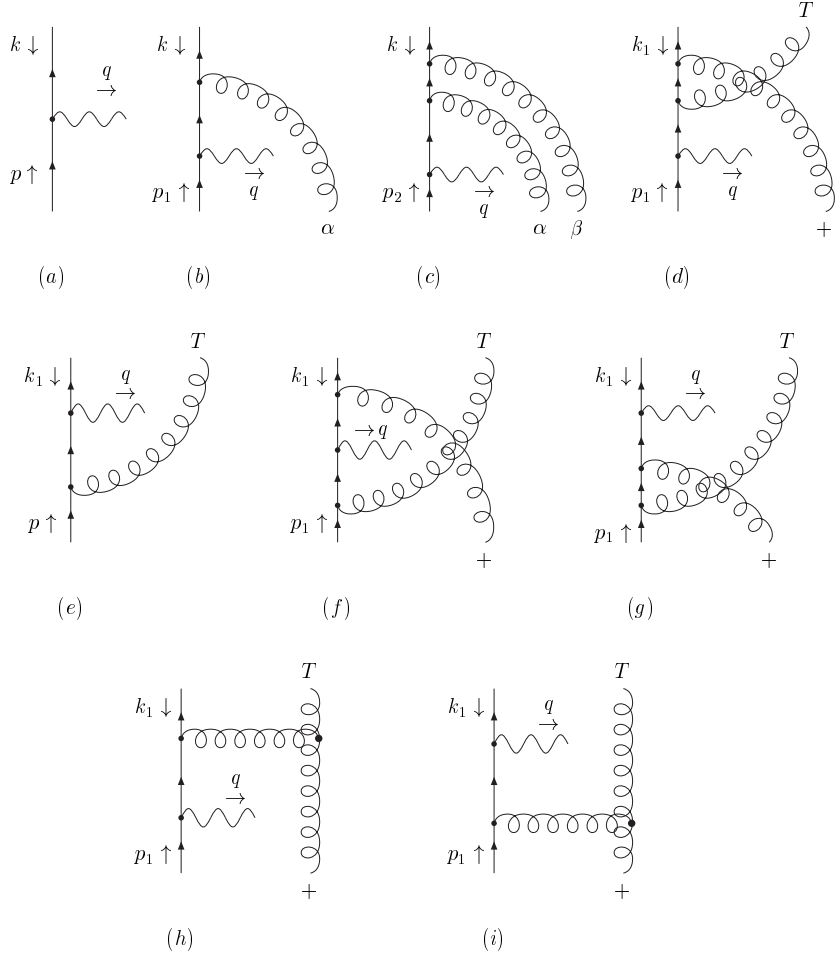}
\caption{\label{links}A subset of diagrams to be considered upon 
inclusion of zero or one $A^+$ gluon(s).}
\end{center}
\end{figure}

To discuss the expressions for the diagrams as given in the figure, we will
introduce the shorthand notation:
\beq
\dbrl \, \ldots \, \dbrr
\equiv \int d^4k\, d^4p \, \delta^4(p+k-q) \, 
\Tr \left[ \ldots \right]
\eeq
Also, we need the precise definitions of $\Phi_A^\alpha(p_1,p)$ and
$\Phi_{AA}^{\alpha \beta}(p_2,p_1,p)$ 
($\alpha, \beta$ not restricted to be transverse) and the antiquark
correlation functions $\overline \Phi (k), \overline \Phi_A^\alpha(k_1,k)$: 
\ba
\Phi_A^\alpha(p_1,p) &=& \int \frac{d^4x}{(2\pi)^4}\, 
\frac{d^4y}{(2\pi)^4}\ e^{ip\cdot y}\,
e^{ip_1\cdot (x-y)} \,
\langle P_1, S_1 \vert \overline \psi(0) \,g A^\alpha(y) \, \psi(x) 
\vert P_1, S_1 \rangle,\\[2 mm]
\Phi_{AA}^{\alpha \beta}(p_2,p_1,p)&=& 
\int \frac{d^4x}{(2\pi)^4}\, \frac{d^4y}{(2\pi)^4}\, \frac{d^4z}{(2\pi)^4}\ 
e^{ip\cdot z}\, e^{ip_1\cdot (y-z)}\, e^{ip_2\cdot (x-y) }\,
\langle P_1, S_1 \vert \overline \psi(0) \,g A^\beta(z) \,g A^\alpha(y) \,
\psi(x)  
\vert P_1, S_1 \rangle,\\[2 mm]
\overline \Phi(k) & = & 
\int\frac{d^4x}{(2\pi)^4} \, e^{-i\,k\cdot x} 
\langle P_2,S_2 \vert \psi(x) \overline \psi(0) \vert P_2,S_2 \rangle
,\\[2 mm]
\overline \Phi_A^\alpha(k_1,k) & = & 
\int\frac{d^4x}{(2\pi)^4} \, \frac{d^4y}{(2\pi)^4}\, e^{-i\,k\cdot (x-y)} 
e^{-i\,k_1\cdot y}
\langle P_2,S_2 \vert \psi(x) gA^\alpha(y)
\overline \psi(0) \vert P_2,S_2 \rangle.
\ea
The hadron momenta $P_1,P_2$ (chosen such that $P_1 \cdot P_2 = {\cal
O} (Q^2)$) and photon momentum $q$ are decomposed as 
\begin{eqnarray}   
&& P_1^\mu \equiv \frac{ Q}{x \sqrt{2}}\,n_+^\mu 
+ \frac{x M_1^2}{ Q\sqrt{2}}\,n_-^\mu,\\
&& P_2^\mu \equiv \frac{\bar{x} M_2^2}{ Q\sqrt{2}}\,n_+^\mu
+ \frac{ Q}{\bar{x}\sqrt{2}}\,n_-^\mu,\\
&&q^\mu \equiv \frac{ Q}{\sqrt{2}}\,n_+^\mu 
+ \frac{ Q}{\sqrt{2}}\,n_-^\mu 
+ q_T^\mu,
\end{eqnarray}
for $Q_T^2 \equiv - q_T^2 \ll Q^2$. 
Furthermore, we decompose the parton momenta as ($i=1,2,\ldots$) 
\ba
&&p \equiv \frac{Q}{\sqrt{2}}\,n_+
+ \frac{(p^2 + \bm{p}_T^2)}{Q \sqrt{2}}\,n_- + p_T
, \\[2 mm]
&&p_i \equiv \frac{\xi_i Q}{\sqrt{2}}\,n_+
+ \frac{(p_i^2 + \bm{p}_{iT}^2)}{\xi_i Q \sqrt{2}}\,n_- + p_{iT}
, \\[2 mm]
&&k \equiv \frac{Q}{\sqrt{2}}\,n_-
+ \frac{(k^2 + \bm{k}_T^2)}{Q \sqrt{2}}\,n_+ + k_T
, \\[2 mm]
&&k_i \equiv \frac{\bar \xi_i Q}{\sqrt{2}}\,n_-
+ \frac{(k_i^2 + \bm{k}_{iT}^2)}{\bar \xi_i Q \sqrt{2}}\,n_+ + k_{iT}.
\ea
We find for diagrams (a)-(d): 
\ba
(a) &=& 
\dbrl \Phi(p) \gamma^\mu \overline \Phi(k) \gamma^\nu 
\dbrr,\\[2 mm]
(b) &=& \dbrl 
\Phi_A^\alpha (p) \gamma^\mu \overline \Phi(k) \gamma_{\alpha T} 
\frac{\slsh{n_+}}{\sqrt{2} Q} \gamma^\nu \dbrr 
+\dbrl \int d^4 p_1 \,\Phi_A^+(p_1,p)\, 
\frac{\sqrt{2}}{Q(1-\xi_1+i\epsilon)} \, \gamma^\mu \overline \Phi(k) 
\gamma^\nu \dbrr
\nn \\[2 mm]
&+& \dbrl \int d^4 p_1 \,\Phi_A^+(p_1,p) \gamma^\mu
\overline \Phi(k)\frac{\sqrt{2}}{Q(1-\xi_1+i\epsilon)}
\left\{ \frac{\slsh{n_+}}{\sqrt{2} Q} \left(\slsh{k_T} + \slsh{p_T}
-\slsh{p_{1T}} \right) - P_- \right\} \gamma^\nu \dbrr,
\label{(b)}\\[2 mm]
(c) &=& \dbrl \int d^4 p_2 \, d^4 p_1 \, 
\Phi_{AA}^{\alpha \, +}(p_2,p_1,p) \, \frac{\sqrt{2}}{Q(1-\xi_1+i\epsilon)}
\, \gamma^\mu \overline \Phi(k) \gamma_{\alpha T} 
\frac{\slsh{n_+}}{\sqrt{2} Q} \gamma^\nu \dbrr,
\label{(c)}\\[2 mm]
(d) &=& \dbrl \int d^4 p_1 \,\Phi_A^+ (p_1,p) \gamma^\mu
\left(
\int d^4 k_1 \, \overline \Phi_D^{\beta}(k_1,k)\gamma_{\beta T} +
\overline \Phi(k) \slsh{k_T}  \right) 
\frac{\sqrt{2}}{Q(1-\xi_1+i\epsilon)} 
\frac{\slsh{n_+}}{\sqrt{2} Q} \gamma^\nu \dbrr.
\ea
In the last equation we have used the fact that in the $A^-=0$ gauge no link
operator will be present in $\overline \Phi_A^\beta$. 
We can then apply the equations of motion to $
\int d^4 k_1 \, \overline \Phi_D^{\beta}(k_1,k) \gamma_{\beta T}$, 
which results (in the $A^-=0$ gauge) in a cancellation of 
the term with the $P_-=\slsh{n_-}\slsh{n_+}/2$ projector 
in the expression for diagram (b).
Anticipating the specific link operators 
that will result in the end, we observe that
\ba
\Phi[-\infty,x](p) &\equiv&  
\int \frac{d^4x}{(2\pi)^4}\, e^{ip\cdot x}\, \langle P_1, S_1 \vert \overline
\psi(0) \, {\cal L}^+ [-\infty,x]\, 
\psi(x) \vert P_1, S_1\rangle \nn\\[2 mm]
&=& \Phi(p) + \int d^4 p_1 \,\Phi_A^+ (p_1,p) 
\frac{\sqrt{2}}{Q(1-\xi_1+i\epsilon)} + \ldots,\\[2 mm]
\Phi_A^\alpha[-\infty,x](p) &\equiv& 
\int \frac{d^4x}{(2\pi)^4}\, e^{ip\cdot x}\, 
\langle P_1, S_1 \vert \overline \psi(0) \, {\cal L}^+
[-\infty,x]\, g A^\alpha(x) \, \psi(x) \vert  P_1, S_1 
\rangle
\nn\\[2 mm]
&=&\Phi_A^\alpha (p) + \int d^4 p_2 \, d^4 p_1 \,  
\Phi_{AA}^{\alpha \, +}(p_2,p_1,p)
\frac{\sqrt{2}}{Q(1-\xi_1+i\epsilon)}+ \ldots,
\label{ADellink}
\ea
where the definition of $\Phi_A^\alpha (p)$ is given in 
Eq.\ (\ref{p1integratedPhi}). 
The sum of the four diagrams (a)-(d) therefore results in (to order $g$ in the 
link operator) 
\ba 
(a)+(b)+(c)+(d) &=& 
\dbrl \Phi[-\infty,x](p) \gamma^\mu \overline \Phi(k) \gamma^\nu 
\dbrr + \dbrl \Phi_A^\alpha[-\infty,x](p) 
\gamma^\mu \overline \Phi(k) \gamma_{\alpha T} 
\frac{\slsh{n_+}}{\sqrt{2} Q} \gamma^\nu \dbrr\nn \\[2 mm]
&+& \dbrl \int d^4 p_1 \,\Phi_A^+(p_1,p) \gamma^\mu
\overline \Phi(k)\frac{\sqrt{2}}{Q(1-\xi_1+i\epsilon)}
\frac{\slsh{n_+}}{\sqrt{2} Q} \left(\slsh{p_T}
-\slsh{p_{1T}} \right) \gamma^\nu \dbrr . \label{abcd1}
\ea
By taking into account the derivative of the link operator, the last 
two terms nicely combine into the following expression containing
exclusively the functions $\Phi[-\infty,x](p)$ and  
$\Phi_D^\alpha[-\infty,x](p)$ (the latter is obtained from 
Eq.\ (\ref{ADellink}) by replacing $gA^\alpha$ by $iD^\alpha$):
\ba
(a)+(b)+(c)+(d) &=& 
\dbrl \Phi[-\infty,x](p) \gamma^\mu \overline \Phi(k) \gamma^\nu 
\dbrr\nn \\[2 mm]
&+&\dbrl \left\{\Phi_D^\alpha[-\infty,x](p) - p^\alpha
\, \Phi[-\infty,x](p) \right\}
\gamma^\mu \overline \Phi(k) \gamma_{\alpha T} 
\frac{\slsh{n_+}}{\sqrt{2} Q} \gamma^\nu \dbrr.\label{abcd2}
\ea
The diagrams (e)-(g) also turn out to combine, yielding the (expected) 
expression:
\ba
(e)+(f)+(g) = -
\dbrl \Phi[-\infty,x](p) \gamma^\mu 
\left\{ \gamma_0 \overline \Phi_D^{\beta\dagger} (k) \gamma_0 + 
k^\beta \overline \Phi(k)
\right\} \gamma^\nu \frac{\slsh{n_-}}{\sqrt{2} Q} \gamma_{\beta
T}\dbrr , 
\label{ef}
\ea
where we used that $\int d^4 k_1 \overline \Phi_D^{\beta}(k_1,k) = 
\gamma_0 \overline \Phi_D^{\beta\dagger} (k) \gamma_0$, when $
\overline \Phi_D^{\beta}(k) \equiv 
\int d^4 k_1 \overline \Phi_D^{\beta}(k,k_1)$, cf.\ Ref.\
\cite{Mulders-Tangerman-96}.   
To arrive at Eq.\ (\ref{ef}) we also made use of the fact that to this order 
$\overline \Phi_A^\alpha$ (with $\alpha$ transverse) will
be proportional to $\slsh{n_-}$ and that 
$\Phi_A^+$ will be proportional to $\slsh{n_+}$.

Until now we have ignored the issue of color matrices. It is easy to see that
in diagrams (d), (f) and (g) the color matrices are in the wrong order for
them to be
absorbed into the correlation functions along with the link operators. 
The diagrams with the triple gluon
vertex should be included to cancel the contributions arising from commuting
the two color matrices. In this way diagram (h) will cancel the term arising 
from diagram (d) and diagram (i) that of diagrams (f) and (g). Hence, by
including diagram (h) and (i) one can include the color matrices in the link
operators. 

This concludes the case of including {\em zero or one\/} $A^+$ gluon(s)
(emanating from $\Phi$, connecting to the left). 
Inclusion of {\em two\/} 
$A^+$ gluons is done in a similar fashion, where we note that
from diagram (c) with $\alpha=\beta=+$ one can see that a product of two
theta-functions in the minus component appears, which gives rise to the 
path-ordering. Knowing this, it is a matter of iteratively treating each order
like 
the first order we have discussed just now, by considering
$\left\{\Phi_D^\alpha[-\infty,x](p) - p^\alpha
\, \Phi[-\infty,x](p) \right\}$ as an effective $\Phi_A^\alpha(p)$. 

When the $A^+$ gluons emanating from $\Phi$ connect to the
right side of the diagrams, then one arrives at the link operator
\beq
{\cal L}^+ [0,-\infty]= \left.
{\cal P} \exp\left(ig\int_{-\infty}^{0^-} dy^-\,
A^+(y)\right)\right|_{y^+=0^+, \bm{y}_T=\bm{0}_T}.
\eeq
Inclusion of arbitrary numbers of $A^+$ gluons emanating from $\Phi$
connecting to either side of the diagram, results 
in the appearance of the following color gauge invariant correlation functions
exclusively:
\ba
\Phi_{\text{GI}}(p) &\equiv&  
\int \frac{d^4x}{(2\pi)^4}\, e^{ip\cdot x}\, \langle P_1, S_1 \vert \overline
\psi(0) \, {\cal L}^+ [0,-\infty] \, {\cal L}^+ 
[-\infty,x]
\, \psi(x) \vert P_1, S_1\rangle, \label{PhiGI}\\[2 mm]
\Phi_{D\, \text{GI}}^\alpha(p_1,p) &\equiv& \int \frac{d^4x}{(2\pi)^4}\, 
\frac{d^4y}{(2\pi)^4}\ e^{ip\cdot y}\, e^{ip_1\cdot (x-y)} \, \langle P_1, S_1
\vert \overline \psi(0) \,
{\cal L}^+ [0,-\infty] \nn\\[2 mm]
&& \times \, 
{\cal L}^+ [-\infty,y] \, iD^\alpha(y)\, {\cal L}^+ 
[y,-\infty]
\, {\cal L}^+ [-\infty,x]\, \psi(x) 
\vert P_1, S_1 \rangle.\label{PhiDGI}
\ea
Hence, Eqs.\ (\ref{PhiGI}) and (\ref{PhiDGI}) are the Drell-Yan process 
extentions of 
Eqs.\ (\ref{PhiDIS}) and (\ref{PhiDDIS}) to the case the correlation functions
not only depend on the lightcone momentum fractions. 

So we conclude that at the tree level the complete set of
diagrams with arbitrary numbers of $A^+$ gluons emanating from $\Phi$ 
will result in the appearance of color gauge invariant correlation
functions (containing link operators and covariant derivatives), with
path-ordered exponentials having paths along lightlike directions extending to
lightcone infinity. Not choosing the $A^-=0$ gauge cannot affect 
$\Phi_{\text{GI}}$ or $\Phi_{D\, \text{GI}}^\alpha$ to this order ($1/Q$) 
and hence, these are the correlation
functions that will appear in any gauge. Choosing the $A^+=0$ gauge instead,
will give rise to similar results for $\overline \Phi$ and $\overline
\Phi_D^\alpha$.
The color gauge invariant matrix elements $\overline 
\Phi_{\text{GI}}(k)$ and $\overline \Phi_{D\, \text{GI}}^\alpha(k_1,k)$ 
contain similar links extending to lightcone infinity, but now involving $A^-$
gluon fields:
\ba
\overline \Phi_{\text{GI}}(k) &\equiv& 
\int\frac{d^4x}{(2\pi)^4} \, e^{-i\,k\cdot x} 
\langle P_2,S_2 \vert \psi(x) \, {\cal L}^- 
[x,-\infty]\, {\cal L}^- [-\infty,0]\, 
\overline \psi(0) \vert P_2,S_2 \rangle
,\\[2 mm]
\overline \Phi_{D\, \text{GI}}^\alpha(k_1,k) &\equiv& 
\int\frac{d^4x}{(2\pi)^4} \, \frac{d^4y}{(2\pi)^4}\, e^{-i\,k\cdot (x-y)} 
e^{-i\,k_1\cdot y}
\langle P_2,S_2 \vert \psi(x) \, {\cal L}^- 
[x,-\infty]\nn\\[2 mm]
&& \times \, {\cal L}^- [-\infty,y] \, iD^\alpha(y)\, {\cal L}^- 
[y,-\infty]
\, {\cal L}^- [-\infty,0]\,
\overline \psi(0) \vert P_2,S_2 \rangle,
\ea  
where 
\begin{equation}
{\cal L}^-[x,-\infty] = \left. {\cal P} \exp \left(ig \int_{-\infty}^{x^+} 
dy^+ \, A^-(y) \right)\right|_{y^- = x^-, \bm{y}_T=\bm{x}_T}.
\label{link2DY}
\end{equation}
One thus finds the following color gauge invariant 
expression for the hadron tensor not integrated
over the transverse momentum of the lepton pair,  
including $1/Q$ contributions:
\ba
{\cal W}^{\mu \nu}(P_1,P_2,q) &=& \frac{1}{3} \,\int d^4k\, d^4p \, \delta^4(p+k-q) \, 
\Biggl\{ 
\Tr \left[ \Phi_{\text{GI}}(p) \gamma^\mu \overline \Phi_{\text{GI}}(k) 
\gamma^\nu \right] 
\nn \\[2 mm]
&+& \Tr \left[ \left\{\int d^4 p_1 \, \Phi_{D\, \text{GI}}^\alpha(p_1,p) -
p^\alpha 
\, \Phi_{\text{GI}}(p) \right\} \gamma^\mu \overline \Phi_{\text{GI}}(k) 
\gamma_{\alpha T} 
\frac{\slsh{n_+}}{\sqrt{2} Q} \gamma^\nu \right]
\nn \\[2 mm]
&-&\Tr  \left[ \Phi_{\text{GI}}(p) \gamma^\mu 
\left\{ \int d^4 k_1 \, \overline \Phi_{D\, \text{GI}}^{\beta}(k_1,k) + 
k^\beta \, \overline \Phi_{\text{GI}}(k)
\right\} \gamma^\nu \frac{\slsh{n_-}}{\sqrt{2} Q} \gamma_{\beta T}\right]
\nn \\[2 mm]
&+&
\Tr \left[ \left\{\int d^4 p_1 \, \Phi_{D\, \text{GI}}^\alpha(p,p_1) - p^\alpha
\, \Phi_{\text{GI}}(p) \right\} \gamma^\mu \frac{\slsh{n_+}}{\sqrt{2} Q}
\gamma_{\alpha T}  \overline \Phi_{\text{GI}}(k) \gamma^\nu \right]
\nn \\[2 mm]
&-& \Tr 
\left[ \Phi_{\text{GI}}(p) \gamma_{\beta T}\frac{\slsh{n_-}}{\sqrt{2} Q} 
\gamma^\mu \left\{ \int d^4 k_1 \, 
\overline \Phi_{D\, \text{GI}}^{\beta}(k,k_1) + 
k^\beta \, \overline \Phi_{\text{GI}}(k) \right\} \gamma^\nu \right]\Biggl\}.
\label{Complete}
\ea
This expression 
still contains some order $1/Q^2$ contributions, which we ignore to keep the
notation as simple as possible.
To the order we consider one can for instance replace
\beq
\delta^4(p+k-q) \to \delta(q^+-p^+)\, \delta(q^--k^-)\, 
\delta^2(\bm{p}_T^{}+\bm{k}_T^{}-\bm{q}_T^{}).
\label{deltafn0}
\eeq
To obtain the hadron tensor expression Eq.\ (\ref{Complete}) 
we have 
already used these delta functions, for instance in the parametrizations 
of the momenta $q,p,k$. 

We would like to emphasize that we have not obtained color
gauge invariance at the cost of introducing a path dependence. The expansion
in orders of $1/Q$ does not leave any freedom of choosing the paths for the
leading and next-to-leading twist terms. 

\subsection{\label{IIIA2}Color gauge invariant correlation functions to all
orders in $\alpha_s$}

In order to extend the above tree level analysis to all orders in $\alpha_s$
in the hard scattering part, we will first discuss two limiting cases. The
first case we consider is the cross section integrated over the transverse
momentum $\bqt$ of the lepton pair. The second case is the cross section at
measured $\bqt$ (with $\bm{q}_T^2 \equiv Q_T^2 \ll Q^2$), but restricted to 
the leading twist.  

In case one integrates the cross section, and hence the hadron tensor, 
over the transverse momentum $\bqt$ of the 
lepton pair, then the situation simplifies considerably. After a collinear
expansion of the hard scattering parts, the transverse momenta of the partons
can be integrated over each separately. The hadron tensor
will be expressed in terms of collinear partons only, i.e., they are collinear 
to the momentum direction of the parent hadron (neglecting target mass 
corrections). More specifically, the hadron tensor contains the above gauge 
invariant correlation functions
$\Phi, \Phi_D^\alpha$ ($\overline \Phi$, $\overline \Phi_D^\alpha$) 
integrated over the $-$ ($+$) and transverse 
momentum components. They depend only on the lightcone momentum fractions, 
such as the ones that appear in the DIS cross section, cf.\ 
Eq.\ (\ref{PhiDIS}) and (\ref{PhiDDIS}). In that case one can apply 
Ward identities to gluon fields with polarizations proportional to
their momenta, to prove that the above results are also 
correct to all orders in $\alpha_s$ in the hard scattering part. 
This has been done 
for the leading twist contribution in for instance Refs.\ \cite{Rad-Efr,EGM},
Ref.\ \cite{CSS85-b} (cf.\ its Fig.\ 4.4 and appendix) and 
Ref.\ \cite{CSS89} (cf.\ its Figs.\ 19 and 25 and Eqs.\ (131) and (133)). 
It is completely analogous to the application of Eq.\ (\ref{WardDIS}) in DIS.
Also, it straightforwardly applies to the $1/Q$
corrections for which a factorization theorem has been discussed in Ref.\
\cite{Qiu-Sterman-90}. The same Ward identities can be applied,
since the additional $A_T$ gluon has physical polarization and is on-shell 
and therefore, does not affect how the collinear gluons appear in the Ward 
identities. The latter just produce link operators on
both side of the $A_T$ field and since in this case the links are independent
of transverse coordinates, the replacement $gA_T \to iD_T -i\partial_T$
immediately gives the result for $\int d^2 \bqt \, 
{\cal W}^{\mu\nu}(P_1,P_2,q)$, using ${\cal W}^{\mu\nu}$ from Eq.\ (\ref{Complete}) 
with hard scattering parts $H$ containing all orders in $\alpha_s$: 
\begin{eqnarray}
&&\int d^2 \bqt \, {\cal W}^{\mu\nu} (x, \bar{x}, \bqt) 
= \frac{1}{3} \,\Biggl\{ 
\text{Tr}\left[ \Phi_{\text{GI}} (x) \overline \Phi_{\text{GI}} 
(\bar{x}) H^{\mu\nu}(x,\bar{x}) \right]\nn  \\
&& \quad
+ \text{Tr}\left[ \int\, dy \, \left\{ \Phi_{D\, \text{GI}}^\alpha (y,x) - 
\delta(y-x) \int\, d^4 p \, \delta (p^+-xP_1^+) \, p^\alpha 
\, \Phi_{\text{GI}}(p)  \right\} 
\overline \Phi_{\text{GI}} 
(\bar{x}) \, H_{1\, \alpha}^{\mu\nu}(y,x,\bar{x}) \right]
\nonumber \\[3mm] 
&& \quad +\text{Tr} \left[ \Phi_{\text{GI}} (x) \, \int\, d\bar{y} \,  
\left\{\overline \Phi_{D\, \text{GI}}^\alpha 
(\bar{y},\bar{x}) 
+ \delta(\bar y- \bar x) \int\, d^4 k \, \delta (k^--\bar{x} P_2^-) \,  
k^\alpha \overline \Phi_{\text{GI}}(k) \right\}
H_{2 \, \alpha}^{\mu\nu}(x,\bar{y},\bar{x})  
\right] \Biggr\}.\label{qtWmunu2}
\end{eqnarray}
This is just a schematic way of writing, since the four spinor indices on the
hard scattering part $H^{\mu\nu}$ and $H_\alpha^{\mu\nu}$ connects to both
$\Phi$ and $\overline \Phi$. At tree level this equation is 
Eq.\ (\ref{qtWmunu}) re-expressed in terms of color gauge invariant
functions, Eqs.\ (\ref{PhiDIS}), (\ref{PhiDDIS}) 
and the obvious extensions $\overline
\Phi_{\text{GI}}(\bar x)$ and $\overline \Phi_{D\, \text{GI}}^\alpha (\bar x, 
\bar y)$. This holds in case soft gluon poles are 
assumed to be absent. Our previous tree level analysis is in
this case an explicit demonstration of a result valid to  
all orders in $\alpha_s$, since soft gluons can be shown to cancel
\cite{Qiu-Sterman-90}. 

To apply such Ward identities to the more extended case we considered here, 
namely to the cross 
section differential in the transverse momentum $\bqt$ of the lepton pair 
($Q_T^2 \ll Q^2$), is more 
complicated. The momenta
of the partons are in general not collinear to the parent hadron, since they 
also possess some transverse momentum. Hence, the $A^+$ gluons, i.e., the
gluons that are collinear to the parent hadron do 
not have polarization collinear to their own momenta. The latter type of
gluons are called gluons with longitudinal polarization \cite{CSS89} and are 
the gluons to which Ward identities can be applied directly. 
For the leading twist contribution the
application of such Ward identities still works for $A^+$ gluons, because 
deviations of $A^+$ gluons from longitudinal polarization  
(such deviations are proportional to $A_T$) will be suppressed by inverse 
powers of the hard scale. 
We then find for the leading twist part of Eq.\ (\ref{Complete}) 
with a hard part $H$ 
containing all orders in $\alpha_s$: 
\beq
{\cal W}^{\mu \nu}(x, \bar{x}, \bqt) = \frac{1}{3} \int \, d^2 \bpt\, d^2 \bkt 
\, \delta^2 (\bm{p}_T^{}+\bm{k}_T^{}-\bm{q}_T^{}) \, 
\Tr \left[ \Phi_{\text{GI}}(x,\bpt) \,
\overline \Phi_{\text{GI}}(\bar x, \bkt) \,
H^{\mu\nu}(x,\bar{x},\bpt,\bkt,\bqt) \right]. 
\eeq
However, this is not the complete result, since if one goes beyond the tree 
level, then apart from having a more complicated
hard scattering part, soft gluons have to be resummed into
Sudakov form factors \cite{Col-89}, resulting in a replacement 
\beq
\delta^2(\bm{p}_T^{}+\bm{k}_T^{}-\bm{q}_T^{})\to \int 
\frac{d^2 \bm{b}}{(2\pi)^2} \, e^{-i \bm{b} \cdot
(\bm{p}_T^{}+\bm{k}_T^{}-\bm{q}_T^{})} \, e^{-S(\bm{b})}, 
\label{sudakovreplacement}
\eeq 
where $e^{-S(\bm{b})}$ is the Sudakov form factor. 
This has been shown in Ref.\ \cite{CSS83,CSS85} for the leading twist. 

\vspace{5 mm}
\noindent
For the next-to-leading twist a factorization proof for the cross 
section differential in the transverse momentum of the lepton pair
($Q_T^2 \ll Q^2$) has not 
been given. But we will discuss how Ward identities applied to longitudinally
polarized gluons can be used to show that collinear gluons factorize, i.e., 
they exponentiate into link operators, also at next-to-leading twist (${\cal O}
(1/Q)$).   

From our tree level result we can deduce how the deviations of longitudinally
polarized gluons from collinearity will affect the application of the 
Ward identities. The part of a longitudinally polarized 
gluon field $A_{\text{lp}}^\mu \sim p^\mu$ that deviates from the $n_+$ 
direction (i.e., from its $A^+$ part) will be proportional to its transverse
momentum ($A_{\text{lp}\,T}^\mu \sim p_T^\mu$). The latter part 
will generate derivatives 
of the $A^+$ parts of the other additional longitudinally polarized gluons, 
which to this order {\em can\/} be taken to be collinear. This yields 
the derivative of a link operator that is 
exactly needed to compensate for the effect of the replacement $gA \to 
iD -i\partial$ inside $\Phi_A^\alpha$ (which now contains link operators 
depending on transverse coordinates), in order 
to arrive at the replacement $\Phi_A^\alpha(p) \to 
\Phi_D^\alpha(p) -p^\alpha\Phi(p)$ for correlation functions containing link 
operators. At tree level this is the 
step from Eq.\ (\ref{abcd1}) to Eq.\ (\ref{abcd2}).

To make this a bit more explicit we write down the longitudinally polarized 
gluon field in Fig.\ \ref{dyfig}:
\beq
A_{\text{lp}}^\alpha 
= n_+^\alpha (n_- \cdot A) + \left( p_T^\alpha - p_{1T}^\alpha
\right) \frac{\sqrt{2}}{Q(1-\xi_1+i\epsilon)} (n_- \cdot A) \, \propto \, 
(p^\alpha - p_1^\alpha). 
\label{Ap}
\eeq 
Insertion of this field into the most general hard part (Fig.\ \ref{dyfig}
contains only the tree level hard part) and applying the Ward
identity will result in those diagrams where the longitudinally polarized 
gluon attaches directly to the ``external'' legs of the hard part (those legs
that connect to the soft parts), cf.\ Ref.\ \cite{CSS89}. 
The $A^+$ part of $A_{\text{lp}}^\alpha$ will yield a term in a 
link operator in $\Phi$ (represented by a double line in 
Ref.\ \cite{CSS89}, called an eikonal line) 
and the second term on the r.h.s.\ of Eq.\ (\ref{Ap}) will give rise to the
derivative of a link operator in $\Phi_A^\alpha$. At tree level these are the
first and third term of Eq.\ (\ref{abcd1}), respectively. 

It is easy to verify that this 
will work to all orders in $\alpha_s$ in the hard part, since one has to take
this deviation into account for only one longitudinally polarized gluon. 
Hence, our conclusion is that
transverse momentum dependence in correlation functions 
does not spoil the summation of collinear 
$A^+$ gluons into link operators {\em at next-to-leading twist} (${\cal O}
(1/Q)$). This is 
not in conflict with the known failure of factorization at even 
higher twist \cite{Doria-80}.   

Hence, Eq.\ (\ref{Complete}) can then be written with a hard part $H$ 
containing all orders in $\alpha_s$:
\ba
{\cal W}^{\mu \nu}(P_1,P_2,q) &=& \frac{1}{3} \,\int d^4k\, d^4p \, \delta^4(p+k-q) \, 
\Biggl\{ 
\Tr \left[ \Phi_{\text{GI}}(p) \, \overline \Phi_{\text{GI}}(k) 
\, H^{\mu\nu}(p,k,q) \right] 
\nn \\[2 mm]
&+& \Tr \left[ \int d^4 p_1 \, \left\{\Phi_{D\, \text{GI}}^\alpha(p_1,p) -
\delta(p_1 - p) \, p^\alpha 
\, \Phi_{\text{GI}}(p) \right\} \, \overline \Phi_{\text{GI}}(k) \, 
H_{1\, \alpha}^{\mu\nu}(p,k,q) \right]
\nn \\[2 mm]
&+&\Tr  \left[ \Phi_{\text{GI}}(p) \, \int d^4 k_1 \,  
\left\{ \overline \Phi_{D\, \text{GI}}^{\beta}(k_1,k) + 
\delta(k_1-k) \, k^\beta \, \overline \Phi_{\text{GI}}(k)
\right\} \, H_{2\, \beta}^{\mu\nu}(p,k,q) \right] \Biggl\}.
\label{Complete2}
\ea
Again soft gluons will have to be included and they will modify the delta
function.
 
Application of Ward identities are an essential part of the proof of
factorization. They are used to show that collinear gluons exponentiate, like
discussed above, but also that soft gluons factorize
from the hard part {\em and\/} from the correlation functions. In the
$\bqt$-integrated case the soft gluons must be shown to cancel and in the
case of measured $\bqt$ they must be shown to exponentiate into Sudakov
factors. For the next-to-leading twist the latter case has not been studied 
yet, but the application of collinear Ward identities as discussed above 
will be an essential ingredient. The fact that Ward identities can
be used to show that collinear gluons also exponentiate at next-to-leading 
twist in the case of measured $\bqt$ (with $Q_T^2 \ll Q^2$), 
suggests that collinear Ward identities can also be applied to the soft gluon
exponentiation and a factorization theorem is expected to hold. It could of 
course be the case 
that the replacement Eq.\ (\ref{sudakovreplacement}) is different for these 
subleading terms than for the leading term. This has to be
investigated further. 

Finally, we have to remark that additional divergences are introduced by the 
paths
extending to infinity if one considers $\alpha_s$ corrections connecting to the
paths; these have to be renormalized. 
A discussion on this topic and how to deal with it is given in \cite{CSS83} 
and applies unchanged to the next-to-leading twist results discussed here. 

\subsection{\label{IIIB}Color gauge invariant soft gluon poles}

Until now we have considered the case where we assume all fields at
infinity to vanish inside the hadronic matrix elements. In the case
of soft gluon poles this assumption is relaxed in a specific way. 
To include such contributions
one has to make the following replacement (cf.\ Eq.\ (\ref{qtWmunu})) in the 
first term in 
Eq.\ (\ref{(b)}) for diagram (b) and in Eq.\ (\ref{(c)}) for diagram (c):
\beq
\frac{\slsh{n_+}}{\sqrt{2} Q} \to \frac{\slsh{n_+}}{\sqrt{2} Q} 
\frac{1-\xi_1}{1-\xi_1+i\epsilon}.
\eeq
A soft gluon pole means that the correlation function has a pole at $\xi_1=1$,
i.e., when the gluon has vanishing momentum. This prohibits applying the
inverse of the above replacement \cite{Boer4}. 

If we consider the above replacement the following matrix element (times
$\delta(p^+-p_1^+)$) results (in addition to the matrix elements in Eq.\
(\ref{Complete})): 
\ba
\lefteqn{\int \frac{d^4x}{(2\pi)^4}\, 
\frac{d^4y}{(2\pi)^4}\ e^{ip\cdot y}\, e^{ip_1\cdot (x-y)} \, \langle P_1, S_1
\vert \overline \psi(0) \, {\cal L}^+
[0,-\infty]}\nn \\[2 mm] 
&& \qquad \qquad \qquad \times 
\frac{1}{2} \frac{\partial}{\partial y^-} \left({\cal L}^+ [-\infty,y] \,
gA_T^\alpha(y)\, 
{\cal L}^+ [y,-\infty]\right) \, {\cal L}^+ [-\infty,x]\, \psi(x) 
\vert P_1, S_1 \rangle.
\label{partialA}
\ea
Due to time reversal symmetry and under the assumption of continuity of the
correlation functions, one can show \cite{Boer4} that for soft gluon
poles one has to have antisymmetric boundary conditions on the field $A_T$ at
$y^-=\pm \infty$. In this case one can replace $gA_T \to iD_T$ in the above
matrix element. The identity 
\beq
\left[ \frac{\partial}{\partial y^-}, {\cal L}^+ [-\infty,y] \,
iD^\alpha(y)\, 
{\cal L}^+ [y,-\infty]\right] = 
{\cal L}^+ [-\infty,y] \, gF^{+
\alpha}(y)\, {\cal L}^+ [y,-\infty]
\eeq
is used to arrive from Eq.\ (\ref{partialA}) at 
\ba
\Phi_{F\, \text{GI}}^\alpha(p_1,p) &\equiv& \int \frac{d^4x}{(2\pi)^4}\, 
\frac{d^4y}{(2\pi)^4}\ e^{ip\cdot y}\, e^{ip_1\cdot (x-y)} \, \langle P_1, S_1
\vert \overline \psi(0) \, {\cal L}^+
[0,-\infty]\nn\\[2 mm]
&& \times \, 
{\cal L}^+[-\infty,y] \, \frac{g}{2} F^{+ \alpha}(y)\, {\cal L}^+ 
[y,-\infty]
\, {\cal L}^+ [-\infty,x]\, \psi(x) 
\vert P_1, S_1 \rangle,
\ea
where $\alpha$ is transverse.
The gauge invariant term $\Phi_{F\, \text{GI}}^\alpha(p_1,p)$ will show up in 
the hadron tensor Eq.\ (\ref{Complete}) in exactly the same way as $\Phi_{D\,
\text{GI}}^\alpha(p_1,p)$. In Eq.\ (\ref{Complete}) one can just replace
\beq
\Phi_{D\,
\text{GI}}^\alpha(p_1,p) \to \Phi_{D\,
\text{GI}}^\alpha(p_1,p) + \Phi_{F\, \text{GI}}^\alpha(p_1,p)\, 
\delta(p^+-p_1^+).
\eeq 
We find that upon 
choosing the $A^+=0$ gauge in $\int d^4 p_1 d^4 p\, 
\Phi_{F\, \text{GI}}^\alpha(p_1,p)\delta(p^+-p_1^+)$ that the standard
expression for the soft gluon pole matrix element $T(x,S_T)$ as given in 
\cite{QS-91b,Boer4} results, cf.\ Eqs.\ (\ref{defT}) and (\ref{defPhiF}) (note
that we have now included a factor $g/2$ in the matrix element $\Phi_{F\,
\text{GI}}^\alpha$). 

From the above it is clear that soft gluon poles require nonzero matrix
elements containing $A_T$ fields at $y^-= \pm\infty$ in any gauge. 
On the other hand, we 
have chosen hadronic matrix elements with $A^+$ fields at
$y^-=\pm\infty$ to vanish. The reason is that
otherwise the previously obtained matrix elements will not be color gauge
invariant anymore and their derivation will not even be well-defined, i.e.,
will not result in finite answers, since 
\beq
\int dp_1^+ \Phi_A^+(p_1,p) \frac{\sqrt{2}}{Q(1-\xi_1 + i\epsilon )}
\eeq
will be divergent. 
We simply conclude that soft gluon poles arise from gluon
fields with physical (transverse) polarizations. 

Hence, the inclusion of soft gluon poles can be done in
a color gauge invariant way, even in the case where 
the transverse momentum of the lepton pair is not integrated over. 

As a final point we note that in the color gauge invariant correlation
functions $\Phi_{D\, \text{GI}}^\alpha$ and 
$\Phi_{F\, \text{GI}}^\alpha$ the index $\alpha$ is purely
transverse, reflected in the appearance of $\gamma_T^\alpha$ in Eq.\
(\ref{Complete}) (analogous to the projector $\omega$ in DIS). 
Imposing the additional, but unnecessary constraint of $n_-$-independence 
at this stage will alter the expression for the hadron tensor
and may result in different answers. 

\section{Conclusions}

We have considered the color gauge invariance of a factorized description of 
the Drell-Yan process cross section. The analysis focused on  
next-to-leading twist contributions for polarized scattering (${\cal O}
(1/Q)$) and on the cross section differential in 
the transverse momentum of the lepton pair ($Q_T^2 \ll Q^2$). 
The hadron tensor has been 
expressed in terms of manifestly color gauge invariant, nonlocal operator 
matrix elements. The tree level case was worked out explicitly and Ward
identities, applied to longitudinally polarized gluons, allowed to 
extend the results to all orders in the hard scattering part. This use of Ward
identities in case longitudinally polarized gluons are not equal to collinear
gluons will also be an important ingredient in a full (and still lacking) 
factorization proof for the case under consideration, to be given in a gauge
independent way. The factorization, i.e., exponentiation into Sudakov factors,
of soft gluons is expected to be analogous to the factorization and 
exponentiation of collinear gluons as derived here.  

The leading twist part of the expressions we have derived, are in agreement 
with earlier investigations \cite{CSS83,CSS85}. Moreover, the whole result 
(including the next-to-leading twist) reduces to 
known results \cite{Qiu-Sterman-90} upon integration over the
transverse momentum of the lepton pair. We have also given a color gauge
invariant treatment of soft gluon poles, showing that in any gauge such poles
arise from gluon fields with physical (transverse) polarizations. 
In addition, we have extensively discussed the discrepancy between the 
results of Ref.\
\cite{Hammon-97} and \cite{Boer4} for a single transverse spin asymmetry in 
the Drell-Yan process. 
This asymmetry appears at the tree level in case soft gluon
poles in the twist-three matrix elements are present.  
We have demonstrated that the
requirement of $n_-$-independence of the resulting 
factorized hadron tensor 
expression as imposed in Refs.\ \cite{Hammon-97,Efremov-Teryaev-84} 
is unnecessary from the point of Lorentz and gauge invariance and 
leads to a different answer in the case of the above
mentioned asymmetry. Without imposing this requirement the derivative 
term will not be present in the asymmetry expression {\em at tree level}. 
We have carefully
considered the color gauge invariance of the 
description of the cross section in terms of the matrix elements, 
to conclude that only physical, i.e.,
transverse gluon degrees of freedom contribute and hence, arrive at the 
expression $A_2$ 
(Eq.\ (\ref{BoerA}) or equivalently 
Eq.\ (\ref{BoerA2})) for the asymmetry.  

\acknowledgments 
We would like to thank John Collins, Alex Henneman, Xiangdong Ji, 
Andreas Sch\"afer, Oleg Teryaev and Raju Venugopalan for valuable 
discussions on this subject. D.B.\ also thanks RIKEN, 
Brookhaven National Laboratory and the U.S.\ 
Department of Energy (contract number DE-AC02-98CH10886) for
providing the facilities essential for the completion of this work.

\end{document}